# Optically gated transduction enables silicon electrical readout of mid-infrared light

Shutao Zhang[1, 2, 3], Jiaqing Shen[2, 4], Chongwu Wang[5], Jiahui Xu[6], Wei Chen[2, 4], Qing Yang Steve Wu[7], Yan Liu[2], Xie Yuen Gok[2], Febiana Tjiptoharsono[7], Xiaogang Liu[8], Qi Jie Wang[5,9], Joel K. W. Yang[1, 10, *], Jinfeng Zhu[4, *], Zhaogang Dong[1, 2, *]

[1]Singapore University of Technology and Design, 8 Somapah Road, 487372, Singapore

[2]Quantum Innovation Centre (Q.InC), Agency for Science Technology and Research (A*STAR), 2 Fusionopolis Way, Innovis #08-03, Singapore 138634, Republic of Singapore

[3]Department of Materials Science and Engineering, National University of Singapore, 9 Engineering Drive 1, Singapore 117575

[4]Institute of Electromagnetics and Acoustics and Key Laboratory of Electromagnetic Wave Science and Detection Technology, Xiamen University, Xiamen, Fujian 361005, China

[5]School of Electrical and Electronic Engineering, Nanyang Technological University, Singapore 639798, Singapore.

[6]Institute of Flexible Electronics (IFE, Future Technologies), Xiang'an Campus, Xiamen University, Xiamen, China

[7]Institute of Materials Research and Engineering (IMRE), Agency for Science, Technology and Research (A*STAR), 2 Fusionopolis Way, Innovis #08-03, Singapore 138634, Republic of Singapore

[8]Department of Chemistry, National University of Singapore, Singapore, Singapore.

[9]CINTRA (CNRS-International-NTU-THALES Research Alliance), IRL 3288, Research Techno Plaza, 50 Nanyang Drive, Border X Block, Level 6, Singapore, 637553 Singapore

[10]Singapore-HUJ Alliance for Research and Enterprise (SHARE), The Smart Grippers for Soft Robotics (SGSR) Programme, Campus for Research Excellence and Technological Enterprise (CREATE), Singapore 138602, Singapore

*Correspondence and requests for materials should be addressed to J.K.W.Y. (email: joel_yang@sutd.edu.sg), J.Z. (email: jfzhu@xmu.edu.cn), and Z.D. (email: Zhaogang_dong@sutd.edu.sg).

## ABSTRACT

Silicon is the dominant platform for integrated photonics, yet its wide electronic bandgap of 1.1 eV fundamentally prevents efficient detection of mid-infrared (MIR) photons through direct valence band to conduction band excitation. Existing approaches therefore seek to enhance sub-bandgap absorption or introduce external infrared absorbers. Here we demonstrate a profoundly different detection strategy based on optically gated transduction, in which MIR photons modulate a pre-established excitation state rather than directly generating photocarriers. Our device integrates $Nd^{3+}$-doped lanthanide nanoparticles with an inverse-designed Al/*p*-Si metasurface supporting dual resonances that simultaneously enhances optical gating in the near infrared and resonantly couples incident MIR radiation. The MIR illumination redistributes the $Nd^{3+}$ emission pathways, producing an optical-state change that is converted into an electrical signal through a silicon Schottky junction. The device achieves room-temperature zero-bias detection at 6.8 μm with a modulation responsivity of 103 μA $W^{-1}$ and resolves combined gate and MIR inputs into three electrically distinguishable current states. Rather than extending silicon absorption into the infrared, this work establishes gated optical transduction as a route for reading MIR optical information on a silicon platform, opening opportunities for integrated infrared sensing, signal processing and programmable photonic systems.

## 1. Introduction

Mid-infrared (MIR) light is attractive for encoding optical states because its low-energy photons probe molecular vibrational fingerprints. [1, 2, 3, 4] However, the low photon energy also makes MIR signals difficult to read using silicon optoelectronics. [5, 6, 7] Silicon provides a mature platform for microfabrication, device integration and electrical readout, [8, 9, 10, 11] but MIR photon energy (typically 0.05-0.50 eV) lies far below silicon bandgap (*i.e.*, ~1.1 eV) and cannot efficiently generate band-to-band photocarriers. [12, 13] For integrated infrared systems aimed at coding, state recognition and on-chip signal processing, the desired operation is not only an intensity-dependent photocurrent. It is the reliable mapping of optical input states into distinguishable electrical output states. [14] The central challenge for silicon-based MIR devices should therefore be reframed. Rather than forcing silicon to absorb low-energy MIR photons directly, an effective transduction pathway is needed between the MIR optical state and the silicon electrical readout.

Existing silicon-based infrared detection strategies have mainly pursued sub-bandgap absorption and extrinsic photoresponse.[15, 16, 17] Hyperdoped silicon[18, 19] and metal/silicon Schottky structures[15, 20, 21] can extend the spectral response through deep levels, internal photoemission or hot-carrier processes.[22, 23] In addition, quantum dots,[24, 25, 26, 27] two-dimensional materials[28, 29] and narrow-bandgap chalcogenides[30, 31, 32] can also serve as external infrared absorbers coupled to silicon junctions or readout circuits. These approaches demonstrate that the silicon platform can be extended into the infrared. Yet their operating principle remains absorber-centred: the MIR field is converted into photocarriers or an absorber-mediated signal, which is then collected by silicon electronics. This direct signal-generation paradigm does not inherently provide a means to write, perturb and distinguish electrical states using separate optical inputs. For optical information processing, therefore, the challenge is not simply to increase MIR absorption, but to establish a transduction pathway in which the MIR input can modulate an electrically readable state.

To move beyond absorber-centred detection, MIR information can instead be converted through an intermediate optical transduction process before entering the silicon readout channel. Lanthanide nanoparticles provide such a route because the discrete energy levels of $Nd^{3+}$ generate near-infrared emission at the wavelengths of 806 nm and 866 nm under external excitation, while the relative intensities of these channels can be modified by MIR irradiation.[33, 34] In this scheme, the MIR input redistributes the lanthanide excited-state population and emission branching rather than directly generating photocarriers in silicon.[35] Here, we implement this concept using an optically gated silicon-based MIR meta-detector that integrates $NaYF_4:Nd^{3+}@NaYF_4$ lanthanide doped nanoparticles (LNPs), an inverse-designed dual-band Al/$p$-Si metasurface, and a silicon electrical readout junction.[36, 37, 38, 39, 40] The metasurface enhances the interaction of the LNPs with both the gate and MIR fields and couples the transduced near-infrared emission to the silicon readout. A 740 nm light beam operates as a gate to establish an electrically accessible state, while additional MIR illumination modulates this state through $Nd^{3+}$-mediated luminescence redistribution and resonantly enhanced optical coupling.[41, 42, 43] This process converts a low-energy MIR input into a resolvable current change, enabling room-temperature MIR state recognition and multilevel electrical mapping without requiring direct band-to-band absorption of MIR photons in silicon.

## 2. Results

### 2.1 Optically gated silicon-based MIR state transduction

The device was designed to read MIR optical states through an optically gated transduction pathway rather than through direct MIR absorption in silicon. At the wavelength of 6.8 μm, the photon energy is approximately 0.18 eV, which is substantially below the silicon bandgap and it is therefore insufficient for efficient band-to-band photocarrier generation. The light gate $L_G$ at the wavelength of 740 nm excites the $Nd^{3+}$ ensemble and establishes a baseline photocurrent in silicon. In the presence of this gate, the MIR light $L_M$ at the wavelength of 6.8 μm perturbs the $Nd^{3+}$ excited-state population, modifies its near-infrared emission, and thereby produces an additional differential photocurrent in the junction (Fig. 1a).

The MIR information is first converted into a near-infrared optical output through the $Nd^{3+}$ energy-level system. Under $L_G$ excitation, $Nd^{3+}$ ions populate the metastable ($^4F_{3/2}$) state and emit through the 806 nm and 866 nm channels. MIR absorption promotes the excited ions to higher-lying states, followed by multi-phonon relaxation and redistribution of the emitting-state populations (Fig. 1b). This process enhances the 806 nm emission channel, while depleting the 866 nm emission channel. It therefore changes the relative populations of two fixed emission channels rather than shifting the emission wavelength continuously.

The Al/$p$-Si metasurface provides the optical coupling required for both stages of the transduction process. Its visible/NIR resonance enhances optical-state writing and coupling of the $Nd^{3+}$ emission to the silicon readout, whereas its MIR resonance strengthens the interaction between $L_M$ and the LNP–metasurface system (Fig. 1c). Under a fixed 740-nm excitation power of 1.0 μW and an incident 6.8-μm MIR power of 1 mW at the sample, the $Nd^{3+}$ emission exhibits pronounced spectral redistribution (Fig. 1e). The 806-nm emission increases while the 866-nm emission decreases, raising the emission intensity ratio ($I_{806}/I_{866}$) from 0.09 to 1.02. Integration of the two emission bands reveals that the increase around 806 nm is approximately 3.6 times the magnitude of the decrease around 866 nm. This asymmetric redistribution reflects MIR-induced modulation of the $Nd^{3+}$ excited-state population established by the 740-nm excitation, rather than requiring a substantial increase in the total emitted photon population.

The dual-resonant Al/*p*-Si metasurface is designed to enhance the electrical readout of this spectral redistribution through complementary optical functions. The MIR resonance strengthens the local interaction between the incident MIR field and the LNP-loaded region, facilitating MIR-assisted modulation of the $Nd^{3+}$ emission pathways. Meanwhile, the near-infrared resonance is designed to preferentially enhance silicon absorption of the MIR-increased 806-nm emission relative to the depleted 866-nm channel. Consequently, the redistributed emission can produce a larger net photocarrier contribution in silicon even without a substantial increase in the total emitted photon flux. Under visible-light excitation, the LNP-integrated Al/*p*-Si metasurface exhibits a 16.8-fold enhancement in near-infrared photoluminescence compared with the unpatterned region (Supplementary Fig. S1), demonstrating metasurface-assisted emission enhancement. The MIR-induced spectral redistribution was characterized separately on a transmissive substrate because the Al/*p*-Si device does not support the required MIR transmission geometry. SEM characterization further confirms the integration of $NaYF_4:Nd^{3+}@NaYF_4$ nanoparticles around the patterned nanostructures (Fig. 1d), establishing the structural basis for the coupled optical transduction and silicon electrical readout. [44, 45, 46, 47, 48, 49, 50, 51, 52, 53, 54]

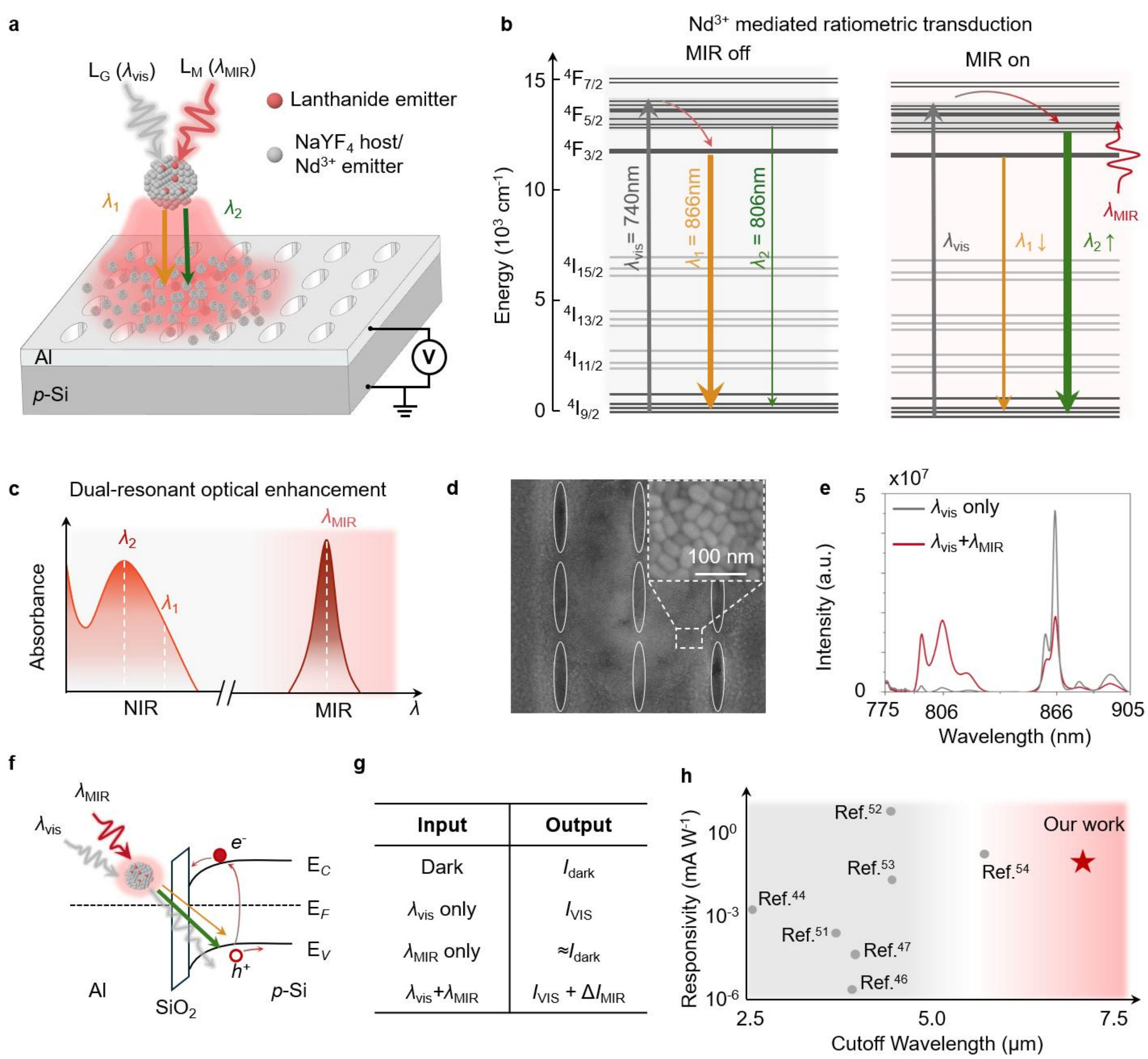

**Figure 1. Optically gated MIR transduction and silicon electrical readout. (a)** Schematic of the Al/*p*-Si metasurface integrated with $NaYF_4:Nd^{3+}@NaYF_4$ core–shell lanthanide nanoparticles, showing $Nd^{3+}$ emitters embedded within the $NaYF_4$ host lattice and their optical coupling to the silicon readout. The light gate $L_G$ writes an electrically readable silicon state and the MIR light incidence $L_M$ modulates the written state. **(b)** $Nd^{3+}$-mediated MIR-to-NIR transduction through excited-state absorption and multi-phonon relaxation. **(c)** Dual-band resonant response of the Al/*p*-Si metasurface for optical gating and MIR modulation. **(d)** SEM image of the LNP-integrated Al/*p*-Si metasurface, showing $NaYF_4:Nd^{3+}@NaYF_4$ LNPs distributed around the nanostructures. **(e)** Photoluminescence spectra of the $Nd^{3+}$-doped LNPs under fixed 740 nm excitation with the 6.8 μm MIR beam off and on, showing an MIR-induced enhancement of the 806 nm channel and suppression of the 866 nm channel. **(f), (g)** Band diagram and optical-input-to-current mapping of the Al/p-Si junction. $L_G$ and $L_M$ denote the optical gate and MIR signal, respectively; $\lambda_1$ and $\lambda_2$ represent the 866- and 806-nm emission channels; $E_C$, $E_F$ and $E_F$ denote the conduction-band edge, Fermi level and valence-band edge, respectively; and $e^-$ and $h^+$ represent electrons and holes. **(h)** Responsivity comparison with other miniaturized silicon-based infrared detectors.

The transduced near-infrared photons are subsequently converted into an electrical signal through the silicon junction readout. As illustrated in Fig. 1f, the gate beam and the LNP emission generate minority carriers in p-Si. These carriers are separated and collected by the junction, producing an electrically readable current. The electrical carrier-collection mechanism is further discussed in Supplementary Note 2. The optical inputs are therefore mapped into distinct current states (Fig. 1g): the dark condition produces $I_{\mathrm{dark}}$; $L_G$ establishes the gate-written current $I_G$; $L_M$ alone remains close to the dark state; and simultaneous $L_G$ and $L_M$ illumination produces $I_G + \Delta I_{\mathrm{MIR}}$.

This gate-dependent mapping enables room-temperature silicon electrical readout at 6.8 μm despite the sub-bandgap MIR photon energy. Comparison with state-of-art miniaturized silicon-based infrared detectors highlights the extended operating wavelength achieved through the combined effects of $Nd^{3+}$ transduction, dual-band resonant enhancement, and silicon junction readout (Fig. 1h).

## 2.2 Dual-band Si-Al metasurface design for gate writing and MIR modulation

To integrate visible light gating and MIR modulation within the same silicon Schottky device, the metasurface must control local optical fields in two spectrally separated bands. The near-infrared resonance is designed to preferentially enhance the effective silicon absorption of the 806-nm emission relative to the 866-nm channel, thereby maximizing the MIR-induced differential photocurrent. In parallel, the MIR resonance enhances the optical interaction between the incident 6.8-μm field and the LNP-loaded region, supporting the MIR-induced redistribution of $Nd^{3+}$ emission. Fig. 2a illustrates the Particle Swarm Optimization (PSO)-based inverse-design workflow. The resonances were positioned to enhance optical interaction with the LNPs in the pump band, favor Si absorption of the 806-nm emission over the 866-nm channel and strengthen coupling to the MIR input. The geometrical parameters, target spectra, and initial structure distribution are used as inputs, and parallel particle searches are performed to minimize the loss function $\xi$:

$$\xi(x) = \omega_{NIR}|A(\lambda_{\mathrm{NIR}}) - A_1|^2 + \omega_{MIR}|A(\lambda_{\mathrm{MIR}}) - A_2|^2 \quad (1)$$

where $A(\lambda_{\mathrm{NIR}})$ and $A(\lambda_{\mathrm{MIR}})$ are the absorption responses in the NIR and MIR target bands, respectively, and $\omega_{NIR}$ and $\omega_{MIR}$ are weighting factors. The particle-wise loss values progressively converge during the optimization, while the global-best loss decreases and subsequently stabilizes with increasing iteration number (Fig. 2b, c). This convergence indicates that the optimization framework identifies geometries satisfying the dual-band spectral objective under the imposed structural constraints.

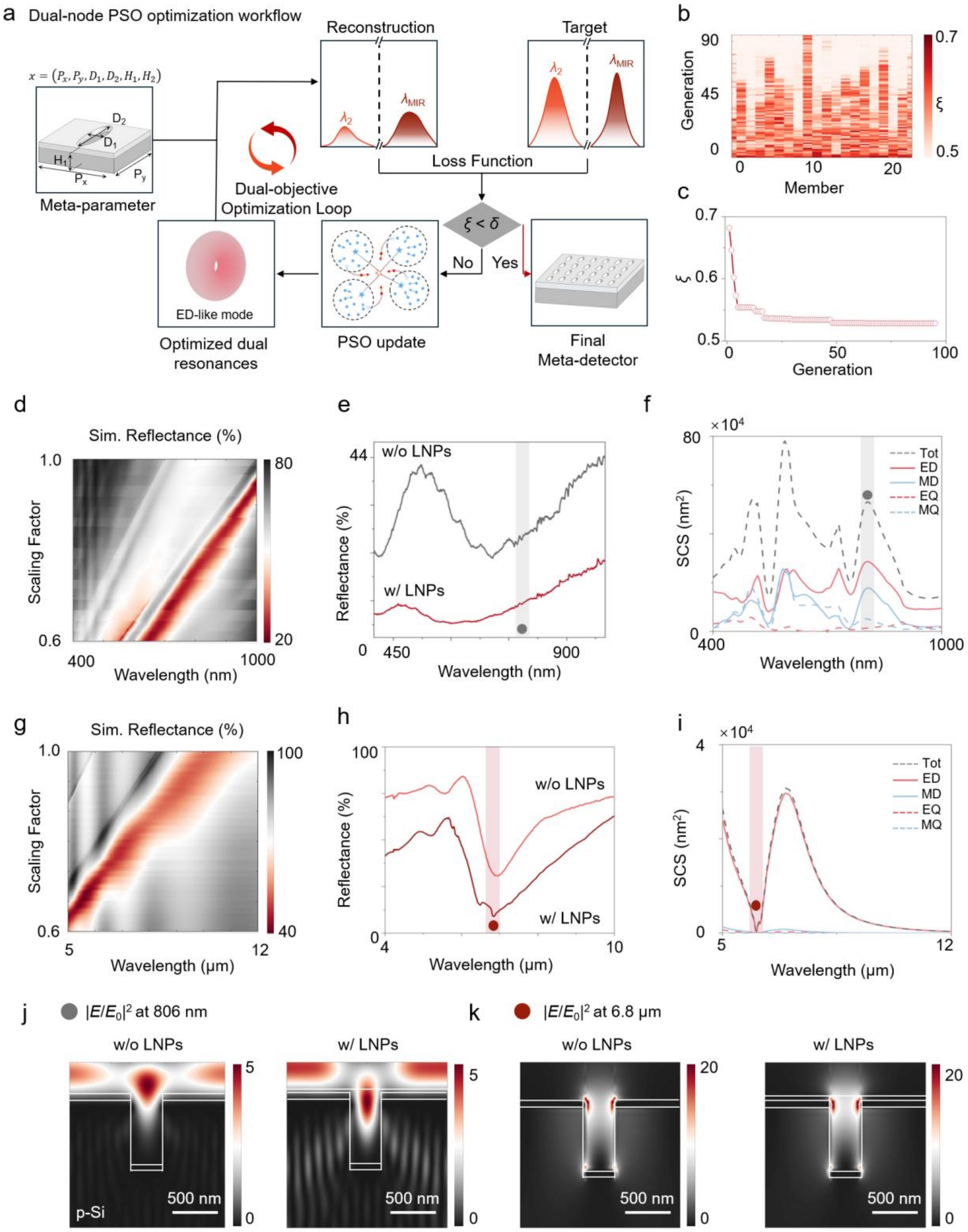

**Figure 2. Physics-inspired inverse design and field coupling of the dual-band Si-Al metasurface. (a)** Particle swarm optimization-based inverse design workflow for obtaining a Si-Al metasurface matched both NIR gating and MIR modulation. (**b**), (**c**) Evolution of the particle-wise loss $\xi$ distribution and global-best loss during the PSO process. (**d**) Simulated reflection spectra in the visible/NIR band. Geometrical scaling

tunes the reflection dip across 700-900 nm, matching the gate beam and $Nd^{3+}$ emission channels. (**e**) Measured visible/NIR reflection spectra with and without LNP integration. Reduced reflection after LNP loading indicates coupling between the LNPs and the metasurface field. (**f**) Multipole decomposition of the selected device in the visible/near-infrared band. (**g**) Simulated MIR reflection spectra showing a tunable resonance in the MIR signal band. (**h**) MIR reflection spectra with and without LNP integration, showing enhanced MIR interaction and field dissipation after LNP loading. (**i**) Multipole decomposition of the selected device in the MIR band. (**j**), (**k**) Near-field intensity distributions $|E/E_0|^2$ at 806 nm and in the 6.8 μm, with and without LNPs, showing that the two resonances couple respectively to the Schottky-active region and the LNP-loaded region.

In the visible/NIR region, simulated reflection spectra show that geometrical scaling tunes the reflection dip across the wavelength range of 700-900 nm, with a calculated absorption of approximately 90%, matching the gate beam and the $Nd^{3+}$ emission channels (Fig. 2d). The measured reflection spectra follow the simulated trend and show a characteristic response near 806 nm; after LNP integration, reflection in the resonant region decreases by approximately 10%, indicating optical coupling between the LNPs and the localized field of the Al/*p*-Si metasurface (Fig. 2e). Multipole decomposition shows that the visible/NIR response is predominantly governed by the magnetic-dipole contribution, confirming that the spectral feature originates from a resonant optical mode rather than a non-resonant absorption background (Fig. 2f). In the MIR region, the metasurface exhibits a scaling-dependent reflection feature over the wavelength range of 5.6-7.4 μm, with a calculated absorption of approximately 80% (Fig. 2g). The experimental spectra verify the MIR resonance and its tunability; LNP loading further reduces the reflection by approximately 15%, suggesting that the nanoparticles modify the local dielectric environment and enhance MIR field dissipation (Fig. 2h). The experimentally measured resonances show a systematic redshift with increasing geometric scaling (Supplementary Fig. S2). This scaling-dependent tuning is retained after LNP integration (Supplementary Fig. S3). Multipole analysis further shows that the MIR response also originates from an electrical dipole (*ED*) mode (Fig. 2i). The corresponding simulated response of the metasurface without LNP loading is provided in Supplementary Fig. S4.

The near-field distributions reveal the complementary spatial functions of the two resonances. At 806 nm, the resonant field is localized near the nanogaps and the Si interface (Fig. 2j), providing spatial overlap with the silicon readout region. The intended function of this resonance is to promote silicon absorption of the MIR-enhanced 806-nm emission relative to the 866-nm channel. At 6.8 μm, the field is concentrated near the metallic features and LNP-loaded region (Fig. 2k), supporting enhanced local interaction with the MIR signal. The inverse-designed metasurface therefore provides simultaneous spectral matching and spatial field localization for optical gating, MIR transduction, and subsequent electrical readout.

### 2.3 Room-temperature electrical readout of the optically gated MIR-modulated state

After establishing the dual-band optical response and $Nd^{3+}$-mediated ratiometric transduction, we examined whether the MIR-modulated optical state could be converted into an electrical output at room temperature. Fig. 3a shows the photocurrent measurement setup. The polarization-controlled gate beam $L_G$ and MIR signal beam $L_M$ were combined and focused onto the LNP-integrated Al/$p$-Si metasurface, while the current was recorded in real time using a source meter. An optical photograph of the fabricated device mounted on its electrical-contact carrier is provided in Supplementary Fig. S5. Four optical states dark, $L_G$ only, $L_M$ only, and $L_G + L_M$ were measured to determine whether electrical MIR readout requires the optically gated state. Fig. 3b shows the corresponding current–voltage characteristics. Under $L_M$-only illumination, the current remained close to the dark level, with a responsivity of approximately 1 μA $W^{-1}$, confirming that MIR illumination alone does not generate an appreciable electrical signal in silicon. By contrast, $L_G$ illumination produced a pronounced current increase, demonstrating that the gate beam writes an electrically readable state in the Si junction. At zero bias, the $L_G$-only responsivity reached 26.0 mA $W^{-1}$. Simultaneous $L_G$ and $L_M$ illumination generated an additional current of 103.335 nA at zero bias. The corresponding MIR modulation responsivity was 103 μA $W^{-1}$. The MIR external quantum efficiency was calculated using

$$EQE = \frac{R_{MIR} hc}{q\lambda_{MIR}} \quad (2)$$

where ($R_{MIR}$) is the gate-dependent MIR modulation responsivity and ($\lambda_{MIR} = 6.8\ \mu m$). This gives a MIR EQE of 0.0188 ‰. The additional current observed under $L_G + L_M$ illumination therefore demonstrates that the MIR input is converted into an electrical output through the $L_G$-established readout state. The corresponding Johnson-noise-limited specific detectivity is estimated to be $8.1 \times 10^7$ Jones; detailed performance calculations are provided in Supplementary Note 3.

The MIR-induced differential current reached a plateau at low reverse bias and remained approximately 130 nA at the bias between 0.1 and 1 V (Fig. 3c). The weak bias dependence indicates that effective carrier collection is already achieved at a relatively low bias (Supplementary Fig. S6). The gate-power dependence was further examined to clarify the formation of the electrically readable state (Fig. 3d). Both the photocurrent and responsivity increase with increasing 740-nm excitation power, with the latter exhibiting a progressively weaker increase at higher powers. Under unsaturated excitation, the direct photocarrier generation in Si and the $Nd^{3+}$-mediated near-infrared emission are expected to scale approximately with the incident gate power and therefore cannot alone explain the increasing responsivity. The observed superlinear photocurrent suggests an additional power-dependent enhancement of the effective electrical conversion efficiency, potentially arising from photoinduced modulation of interfacial charge states, recombination losses and carrier collection across the Al-Si junction. As the junction state progressively stabilizes under stronger illumination, the increase in responsivity becomes less pronounced. Relatively low gate power was therefore selected to establish a readable electrical state while minimizing the gate-generated background that could obscure the MIR-induced modulation.

The $L_G$-written state exhibited a response time of approximately 50 ms, demonstrating rapid optical writing and recovery of the silicon readout channel (Fig. 3e). Control measurements further confirmed the cooperative roles of the LNPs and the metasurface (Fig. 3f). The photocurrent measured from the LNP-integrated nanostructured region was approximately five times that of the nanostructure without LNPs and 43 times that of the LNP-integrated planar region. The enhanced electrical output therefore arises from the combined effects of $Nd^{3+}$ transduction and resonant metasurface coupling to the silicon readout.

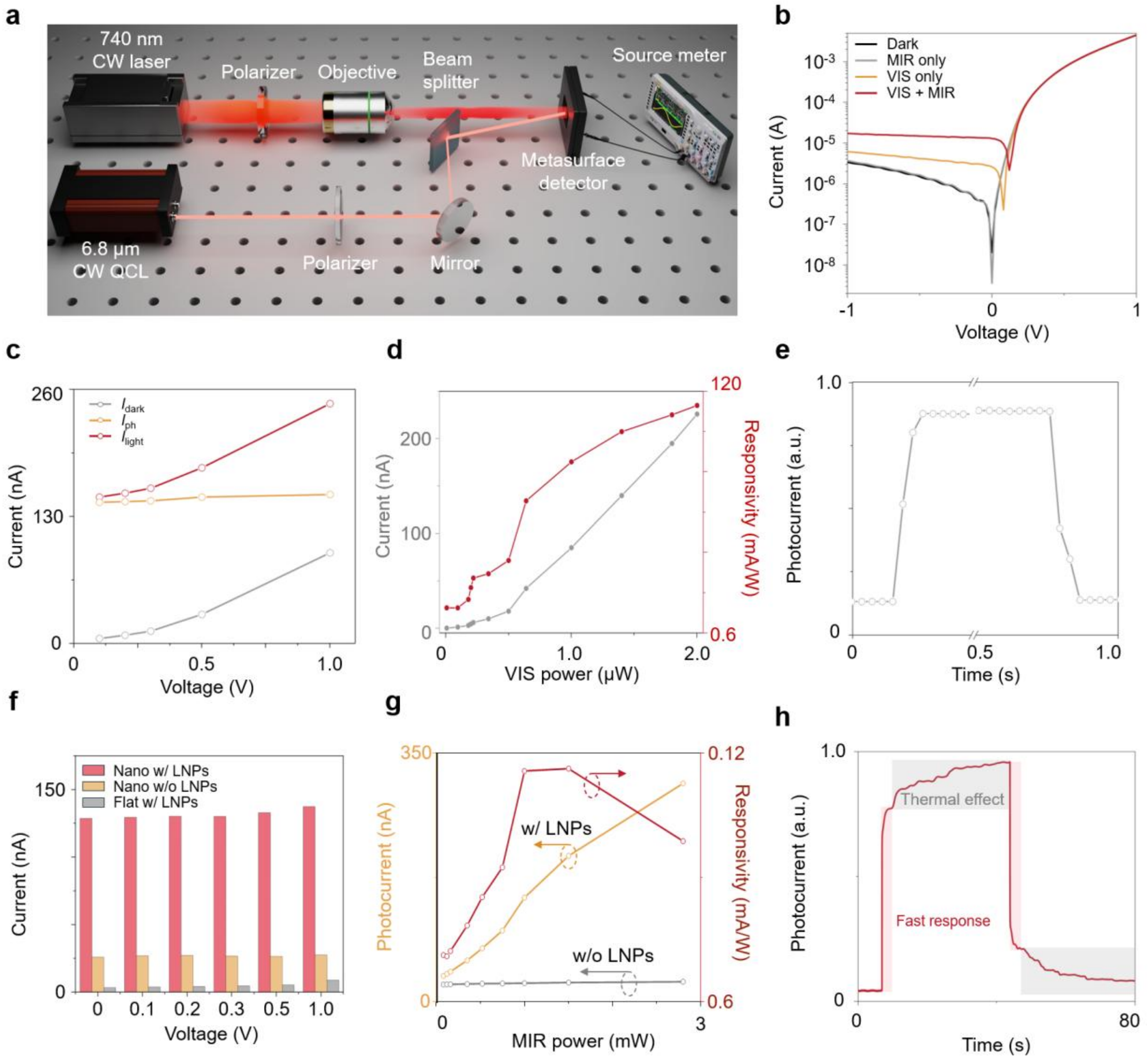

**Figure 3. Room-temperature electrical readout of the optically gated MIR-modulated state. (a)** Photocurrent measurement setup for the LNP-integrated Al/*p*-Si metasurface. **(b)** Current–voltage characteristics under dark, $L_G$-only, $L_M$-only, and $L_G + L_M$ illumination. **(c)** Bias-dependent dark current, illuminated current, and differential photocurrent. **(d)** Gate-power-dependent photocurrent and responsivity. **(e)** Transient writing and recovery of the $L_G$-gated conductive state. **(f)** Photocurrent comparison among different device regions and LNP-integration conditions. **(g)** MIR-power-dependent photocurrent and responsivity under fixed $L_G$ illumination. **(h)** MIR-modulated transient response under fixed $L_G$ illumination, resolved into fast optical-transduction and slow photothermal components.

Under fixed $L_G$ illumination, the MIR-induced differential photocurrent increases monotonically with MIR power (Fig. 3g), whereas MIR-only illumination produces a negligible electrical response over the same power range (Supplementary Fig. S7). This gate-dependent behaviour is consistent with MIR-induced modulation of the pre-excited $Nd^{3+}$ population and its subsequent electrical readout through silicon. At lower MIR powers, the approximately linear increase in photocurrent is consistent with an increasing rate of MIR-assisted excited-state transitions, while the reduced responsivity at higher powers indicates a departure from linear scaling. A relatively high MIR power was therefore selected for transient characterization to obtain a sufficient modulation amplitude.

The MIR-modulated transient exhibits two distinct temporal components (Fig. 3h). The faster component shows a rise time of 90.1 ms and a recovery time of 50.9 ms, with corresponding current amplitudes of 30.9 and 35.5 nA. This response is consistent with the rapid electrical readout of MIR-induced $Nd^{3+}$ emission redistribution through the optically gated Si junction, although its measured timescale may also reflect junction transport and measurement bandwidth. [55] A substantially slower component, characterized by a timescale of approximately 38 s and rise and recovery amplitudes of 19.2 and 12.6 nA, respectively, is consistent with thermally assisted modulation of junction transport, potentially involving interfacial charge redistribution and barrier relaxation. The measured MIR photocurrent therefore comprises a faster gate-dependent modulation and a slower junction-relaxation contribution. [36, 56, 57, 58, 59, 60] Importantly, the temporal separation supports the presence of distinct dynamical processes but does not, by itself, establish an exclusively non-thermal origin of the faster component. The corresponding kinetic analysis is provided in Supplementary Note 2.

## 2.4 Multilevel electrical mapping of optically gated MIR signals

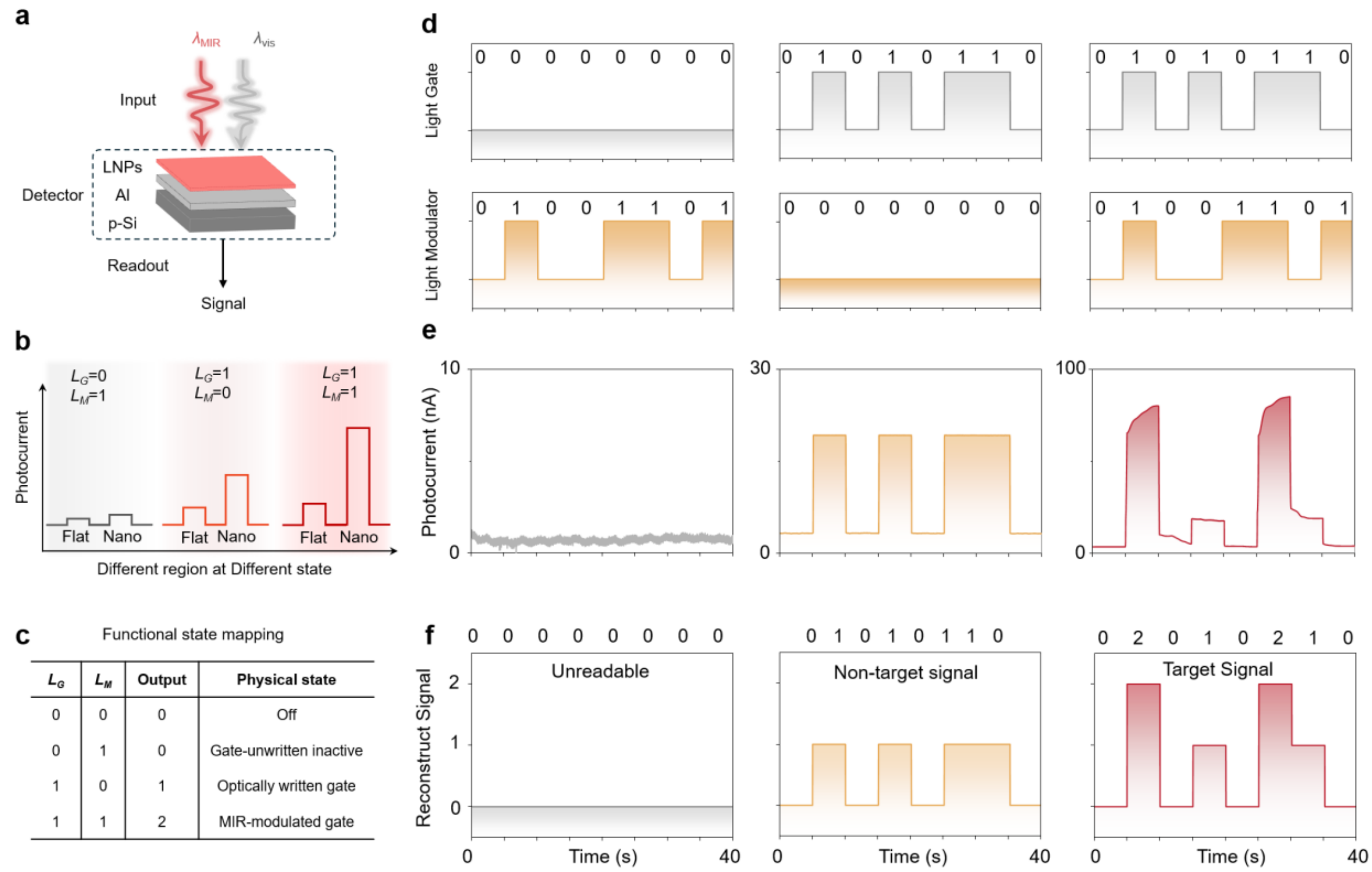

**Figure 4. Multilevel electrical mapping of optically gated MIR signals.** **(a)** Schematic of optically gated MIR signal readout. $L_G$ and $L_M$ are incident on the meta-detector and are converted into different current outputs. (**b)** Mapping between optical inputs and device states: no light, $L_G$ only, and $L_G + L_M$ correspond to the unwritten state, gated conductive state, and MIR-modulated state, respectively. (**c)** Optical-gating electrical readout logic. $L_G$ writes a readable channel, and $L_M$ produces an additional current modulation on the gated state. (d) Coded optical inputs for three configurations: MIR only, gate only and simultaneous gate–MIR input. The gate and MIR sequences are 00000000/01001101, 01010110/00000000 and 01010110/01001101, respectively. (e) Corresponding photocurrent responses. (f) Decoded electrical outputs of 00000000, 01010110 and 02010210, corresponding to unreadable, non-target and target signals, respectively.

Finally, we use the gated readout mechanism to map combined $L_G$and $L_M$inputs into electrical states. Fig. 4a illustrates the optically gated MIR readout process. The LNPs convert the MIR perturbation into a modulation of the near-infrared emission channels, while the silicon junction converts the resulting optical state into an electrical current.

The operating logic consists of three states (Fig. 4b, c). With no optical input, the device remains in an unwritten state and outputs the dark current. With $L_G$ alone, the device enters a gated conductive state and the current rises to $I_{\mathrm{VIS}}$. With simultaneous $L_G$ and $L_M$, MIR light adds a modulation on the gated state and increases the output to $I_{\mathrm{VIS}} + \Delta I_{\mathrm{MIR}}$. Because $L_M$ cannot independently establish an electrically readable silicon state, the device does not read an isolated MIR state; it reads a combined optical state defined by $L_G$ writing and $L_M$ modulation.

Three input configurations were used to examine the electrical mapping of the coded optical signals (Fig. 4d). With the gate off (00000000) and only the MIR sequence (01001101, "M") applied, the photocurrent remains close to the dark level and is decoded as 00000000, indicating that the MIR signal alone is unreadable through the silicon electrical channel. With only the gate sequence (01010110, "V") applied, a binary photocurrent response is obtained and decoded as 01010110, representing a non-target signal. When the gate and MIR sequences are applied together, the MIR modulation introduces an additional current level, giving the multilevel output 02010210, which is identified as the target signal (Fig. 4e, f). These results show that the LNP-integrated Al/$p$-Si metasurface converts combined gate and MIR inputs into distinguishable multilevel electrical states, enabling silicon-compatible readout and processing of MIR-coded signals.

## 3. Conclusions

In conclusion, we demonstrate an optically gated transduction strategy for electrical detection of 6.8 μm mid-infrared (MIR) radiation on a silicon-compatible platform without direct MIR photocarrier generation. By integrating $Nd^{3+}$-doped lanthanide nanoparticles with an inverse-designed dual-resonant Al/$p$-Si metasurface, MIR-induced redistribution of the 806 and 866 nm emission channels is coupled to a visible-light-gated electrical readout. The device achieves a zero-bias MIR-induced current modulation of 103.3 nA, corresponding to a modulation responsivity of 103.3 μA $W^{-1}$ and a MIR-referenced effective external quantum efficiency of 0.0188‰, with a 43-fold photocurrent enhancement over an LNP-coated planar reference. Crucially, MIR illumination modulates a pre-established electrically readable state, enabling three distinguishable outputs corresponding to dark, optically gated, and MIR-modulated conditions. Transient measurements reveal a faster response component with rise and recovery times of 90.1 and 50.9 ms, respectively, alongside slower relaxation associated with additional thermal or junction-state dynamics. These findings establish optically gated state transduction as an alternative to direct MIR photocarrier generation, offering a route toward silicon-compatible infrared sensing and multistate optoelectronic readout.

## 4. Methods

### 4.1 Numerical simulations

The simulations of reflectance spectra and electric field distribution were carried out using the Finite-Difference Time Domain (FDTD) Solutions, ANSYS Inc. software. The refractive index of aluminium (Al) and silicon (Si) were sourced from the Handbook of Chemistry and Physics, and the periodic boundary conditions were applied for the periodic unit cells. The size of the z-direction unit cell in simulation was the same as the x-direction.

**4.2 Fabrication of the metasurface**

The etching mask was patterned on a single-crystalline p-type silicon wafer (10–50 Ω·cm, Si Valley Microelectronics Inc.) using ZEP resist. The resist was spin-coated at 1000 rpm, followed by sequential baking at 180 °C for 2 min and 120 °C for 2 min. Electron beam lithography was performed using an Elionix ELS-7000 system operated at an acceleration voltage of 100 keV and a beam current of 500 pA. After exposure, the sample was developed in o-xylene for 30 s, rinsed with isopropanol, and dried under a nitrogen flow. Silicon etching was then carried out to a depth of 400 nm by inductively coupled plasma etching using the Oxford Instruments Plasmalab System 100. After etching, the residual ZEP mask was removed by soaking the sample in N-methyl-2-pyrrolidone at 85 °C for 2 h, followed by solvent cleaning. Finally, a 30 nm-thick aluminum layer was deposited onto the silicon nanostructures by electron-beam evaporation.

**4.3 Preparation of lanthanide nanoparticles**

$Nd^{3+}$-doped nanocrystals were synthesized using a modified co-precipitation method. Initially, an aqueous solution (4 ml) of Nd $(CH_3CO_2)_3$ and $Y(CH_3CO_2)_3$, in molar ratios of $x/100$ - $x$ (where $x = 0.5$), was mixed with 6.0 ml of oleic acid and 14 ml of 1-octadecene in a 50 ml flask. The mixture was heated to 150 °C for 60 minutes to form lanthanide–oleate coordination complexes. After cooling to 50 °C, a methanol solution (12 ml) containing $NH_4F$ (3.2 mmol) and NaOH (2.0 mmol) were added while stirring for another 60 minutes. The solution was then heated to 100 °C under vacuum for 20 minutes to eliminate volatile residues. Subsequently, the mixture was heated to 290 °C and maintained for 3 hours in an inert nitrogen atmosphere. After cooling to room temperature, the nanocrystals were collected via centrifugation, washed multiple times with ethanol, and dispersed in 4 ml of cyclohexane. To synthesize neodymium-activated core–shell nanocrystals, a shell precursor containing 0.8 mmol $Y(CH_3CO_2)_3$ was prepared similarly. The core nanocrystals were added to this precursor and heated at 80 °C for 40 minutes. Upon cooling to 50 °C, a methanol solution (12 ml) of $NH_4F$ (3.2 mmol) and NaOH (2 mmol) was introduced, followed by 30 minutes of stirring at 50 °C. The reaction was then heated to 100 °C, and impurities were removed before further heating to 290 °C for 2 hours. The core–shell nanocrystals were washed with ethanol and dispersed in cyclohexane. For ligand removal, the nanocrystals were precipitated by adding ethanol, centrifuged, and redispersed in ethanol. The $NaYF_4:Nd^{3+}$@$NaYF_4$ nanoparticles exhibit a predominantly ellipsoidal morphology, with preliminary SEM estimates of approximately 39 nm and 26 nm along the major and minor axes, respectively. Finally, lanthanide nanoparticle films were prepared via drop-casting 20 µl of a ligand-free LNPs solution (10 mg/ml) onto metasurfaces.

### 4.4 Spectroscopic characterization

Optical reflectance spectra in the visible and near-infrared regions were measured using a UV–VIS–NIR microspectrophotometer (CRAIC, QDI-2010) coupled to a polarized broadband illumination source and a 5× objective lens (NA = 0.12). Absolute reflectance was calibrated using a NISR standard reference sample from CRAIC Technologies. For each measurement, spectra were recorded with an integration time of 100 ms and averaged over 50 acquisitions. The morphology of the fabricated metasurfaces was characterized by scanning electron microscopy using a Hitachi SU8220 microscope operated at an acceleration voltage of 1 keV. Mid-infrared reflectance spectra were collected in reflection mode using a microscope-coupled Fourier-transform infrared spectroscopy system (FTIR, Bruker Hyperion 2000). A globar source was used to provide broadband infrared illumination, and the reflected MIR signal was detected by a liquid-nitrogen-cooled mercury cadmium telluride (MCT) detector.

### 4.5 Photoluminescence measurements

Photoluminescence measurements were performed using a customized Olympus BX51 microscope. A 740 nm laser was used to excite the $Nd^{3+}$-doped LNPs, while a tunable MIR laser operated at 6.8 µm was used to modulate the emission. The 740 nm beam was focused onto the sample through a 100× objective lens (NA = 0.9), producing an excitation spot of approximately 15 µm. The emitted light was collected through the same objective, passed through a 740 nm long-pass filter to suppress the excitation beam, and recorded using an Ocean Optics QR Pro spectrometer. Spectra were measured under MIR-off and MIR-on conditions at three randomly selected regions of each sample.

### 4.6 Photocurrent measurements

A multimode laser operating at 740 nm (CNI Laser, MDL-III) was used as the optical gate source. The beam was directed by a 760 nm dichroic mirror and focused onto the LNP-integrated metasurface using a 50× objective lens (Mitutoyo, NA = 0.55). A tunable quantum cascade laser (Daylight Solutions, MIRcat) operated at 6.8 μm was used as the MIR source and focused onto the device using a ZnSe lens with a focal length of 5 cm. The gate and MIR beams were spatially overlapped on the active metasurface region. Current–voltage and transient photocurrent measurements were performed using a computer-controlled dual-channel source meter (Keysight B2912A), which was used to apply the electrical bias and record the device current.

**Notes**

The authors declare no competing financial interests.

**Author contributions**

Z.D. conceived the concepts and supervised the project. Z.D., J.Z. and J.K.W.Y. conceived the concept of a Si-based infrared detector. S.Z., F.T., and Y.L. prepared the metasurface samples. J.X. prepared the conversion nanoparticles samples. S.Z. and X.Y.G., did the SEM characterizations. Q.Y.S.W., and C.W. did the mid-IR device characterizations and interpretations. S.Z., Y.L. and W.C. did numerical simulations and interpretations. S.Z., and J.S. did the optical reflectance measurements. Q.J.W., X.L., and J.D. participated in the discussions and provided the suggestions. The paper was drafted by S.Z. with input from Z.D., J.K.W.Y., J.S., and J.Z. All authors analyzed the data and read and corrected the manuscript before the submission. S.Z., J.S., and C.W. are equal contributions.

**Acknowledgments**

Z.D. and J.K.W.Y. would like to acknowledge the funding support from The National Research Foundation (NRF), Singapore via Grant No. NRF-CRP30-2023-0003. In addition, Z.D. would like to acknowledge the funding support from the Agency for Science, Technology and Research (A*STAR) under its RIE2025 Japan–Singapore Joint Call for Proposals on Quantum 2025 (Project No. H25-MRO3491), and SUTD Kickstarter Initiative (SKI) grant with the award No. SKI 2021_06_05. J.Z. would like to acknowledge the funding support from NSFC (62575250, 62175205) and the Youth Talent Support Program of Fujian Province (Eyas Plan of Fujian Province) [2022]. J.K.W.Y. would like to acknowledge the funding support from National Research Foundation, Prime Minister's Office, Singapore under its Campus of Research Excellence and Technological Enterprise (CREATE) programme, National Research Funding (NRF) Singapore NRF-CRP20-2017-0001 and NRF-NRFI06-2020-0005. Y.L. would like to acknowledge the funding support from A*STAR under the Career Development Award grant (Project No. C243512003). Q.Y.S.W. would like to acknowledge the funding support from The National Research Foundation, Singapore via Grant No. NRF-CRP26-2021-0004 and NRF-NRFI11-2026-001.

## References

1. Tittl A, Michel AKU, Schäferling M, Yin X, Gholipour B, Cui L*, et al.* A switchable mid-infrared plasmonic perfect absorber with multispectral thermal imaging capability. *Advanced Materials* 2015, **27**(31)**:** 4597–4603.

2. Liu X, Zhong Y, Wang Z, Huang T, Lin S, Zou J*, et al.* Electrically programmable pixelated coherent mid-infrared thermal emission. *Nature Communications* 2025, **16**(1)**:** 1665.

3. Yu N, Aieta F, Genevet P, Kats MA, Gaburro Z, Capasso F. A broadband, background-free quarter-wave plate based on plasmonic metasurfaces. *Nano letters* 2012, **12**(12)**:** 6328–6333.

4. Park J, Kang J-H, Kim SJ, Liu X, Brongersma ML. Dynamic reflection phase and polarization control in metasurfaces. *Nano letters* 2017, **17**(1)**:** 407–413.

5. Yang X, Xiong Z, Yu H, Zhang X, Gou J, Zeng G*, et al.* Recent Advances in Si-Based Photodiodes for 1–14 μm Infrared Detection. *Advanced Functional Materials* 2026**:** e13499.

6. Hua S, Divita E, Yu S, Peng B, Roques-Carmes C, Su Z*, et al.* An integrated large-scale photonic accelerator with ultralow latency. *Nature* 2025, **640**(8058)**:** 361–367.

7. Hong S, Zhang L, Wu J, Peng Y, Lyu L, Hu Y*, et al.* Multimode-enabled silicon photonic delay lines: break the delay-density limit. *Light: Science & Applications* 2025, **14**(1)**:** 145.

8. Zhang S, Zhang J, Goh WP, Liu Y, Tjiptoharsono F, Lee HYL*, et al.* Reversible electrical switching of nanostructural color pixels. *Nanophotonics* 2023, **12**(8)**:** 1387–1395.

9. Cai M, Xia F. An all-silicon solution. *Nature Photonics* 2024, **18**(9)**:** 890–891.

10. Liu C, Guo J, Yu L, Li J, Zhang M, Li H*, et al.* Silicon/2D-material photodetectors: from near-infrared to mid-infrared. *Light: Science & Applications* 2021, **10**(1)**:** 123.

11. Jalali M, Yu Y, Xu K, Ng RJ, Dong Z, Wang L*, et al.* Stacking of colors in exfoliable plasmonic superlattices. *Nanoscale* 2016, **8**(42)**:** 18228–18234.

12. Wang F, Hu F, Dai M, Zhu S, Sun F, Duan R, *et al.* A two-dimensional mid-infrared optoelectronic retina enabling simultaneous perception and encoding. *nature communications* 2023, **14**(1)**:** 1938.

13. Zhu S-k, Pan T, Tang C-x, Li A-H, Zheng Z-h, Xu Y, *et al.* Upconversion optical entropy encoding for infrared complex-amplitude imaging. *Light: Science & Applications* 2026, **15**(1)**:** 158.

14. Zhang S, An S, Dai M, Wu QYS, Adanan NQ, Zhang J, *et al.* Chalcogenide Metasurfaces Enabling Ultra-Wideband Detectors From Visible to Mid-infrared. *Advanced Science* 2025, **12**(14)**:** 2413858.

15. Chen W, Zhang S, Wang C, Wu Y, Shi X, Shen J, *et al.* Enabling highly efficient infrared silicon photodetectors via disordered metasurfaces with upconversion nanoparticles. *Science Advances* 2025, **11**(43)**:** eadx7783.

16. Poon JK, Govdeli A, Sharma A, Mu X, Chen F-D, Xue T, *et al.* Silicon photonics for the visible and near-infrared spectrum. *Advances in Optics and Photonics* 2024, **16**(1)**:** 1–59.

17. Ho J, Dong Z, Leong HS, Zhang J, Tjiptoharsono F, Daqiqeh Rezaei S, *et al.* Miniaturizing color-sensitive photodetectors via hybrid nanoantennas toward submicrometer dimensions. *Science Advances* 2022, **8**(47)**:** eadd3868.

18. Xiao Y, Deng K, Zhang K, Xu X, Li Q, Zhang T, *et al.* On-Chip Room-Temperature Operated Short-Wavelength-Infrared Si: S Photodetector with a Vertical Junction. *Advanced Functional Materials* 2024, **34**(49)**:** 2409354.

19. Lischke S, Peczek A, Morgan J, Sun K, Steckler D, Yamamoto Y, *et al.* Ultra-fast germanium photodiode with 3-dB bandwidth of 265 GHz. *Nature Photonics* 2021, **15**(12)**:** 925–931.

20. Li X, Deng Z, Li J, Li Y, Guo L, Jiang Y, *et al.* Hybrid nano-scale Au with ITO structure for a high-performance near-infrared silicon-based photodetector with ultralow dark current. *Photonics Research* 2020, **8**(11)**:** 1662–1670.

21. Zhang S, Pan C-F, Fan Y, Shin J, Liu Y, Liu Y, *et al.* Engineering Hybrid Resonances in Nanophotonics. *arXiv preprint arXiv:260520605* 2026.

22. Liu X, Wu S, Cao X, Tian F, Bodepudi SC, Malik M, *et al.* Plasmon resonance-enhanced graphene nanofilm-based dual-band infrared silicon photodetector. *Photonics Research* 2023, **11**(10)**:** 1657–1666.

23. Zhang G, Zhang S, Zheng L, Wu H, Wang B, He Z, *et al.* Localized surface plasmon resonance enables Si-based near-infrared photodetector. *IEEE Transactions on Electron Devices* 2023, **70**(10)**:** 5497–5500.

24. Wang Y, Wu H, Rodà C, Peng L, Taghipour N, Dosil M, *et al.* Shortwave infrared light detection and ranging using silver telluride quantum dots. *Advanced Materials* 2025, **37**(21)**:** 2500977.

25. Zhou W, Zheng L, Ning Z, Cheng X, Wang F, Xu K, *et al.* Silicon: Quantum dot photovoltage triodes. *Nature Communications* 2021, **12**(1)**:** 6696.

26. Wu H, Liu Z, Wang B, Zheng L, Lian S, Zhang J*, et al.* Integration of PbS quantum dots with 3D-graphene for self-powered broadband photodetectors in image sensors. *Acs Photonics* 2024, **11**(3)**:** 1342–1351.

27. Liu G, Zhou W, Gromyko D, Huang D, Dong Z, Liu R*, et al.* Single-photon generation and manipulation in quantum nanophotonics. *Applied Physics Reviews* 2025, **12**(1).

28. Mao J, Zhang B, Shi Y, Wu X, He Y, Wu D*, et al.* Conformal MoS2/silicon nanowire array heterojunction with enhanced light trapping and effective interface passivation for ultraweak infrared light detection. *Advanced Functional Materials* 2022, **32**(11)**:** 2108174.

29. Yin Y, Cao R, Guo J, Liu C, Li J, Feng X*, et al.* High-speed and high-responsivity hybrid silicon/black-phosphorus waveguide photodetectors at 2 μm. *Laser & Photonics Reviews* 2019, **13**(6)**:** 1900032.

30. Wang D, Wang X, Pan Y, Bian J, Liu K, Guo J*, et al.* A High-Speed Image Sensor Based on Large-Area MoTe2/Si Photodiode Arrays. *Advanced Optical Materials* 2024, **12**(22)**:** 2400547.

31. Guo H, Xia Y, Yu Y, Zhou R, Niu H, Mao X*, et al.* High-speed and broadband spectral photodetectors based on β-In2Se3/Si heterojunction. *Materials Science in Semiconductor Processing* 2022, **138:** 106304.

32. Gautam V, Gautam S, Maurya GK, Kandpal K, Singh B, Ganesan R*, et al.* Investigation of RF sputtered, n-Bi2Se3 heterojunction on p-Si for enhanced NIR optoelectronic applications. *Solar Energy Materials and Solar Cells* 2022, **248:** 112028.

33. Kassir AA, Cheignon C, Charbonnière LJ. Exploitation of luminescent lanthanide nanoparticles for a sensitivity-enhanced ELISA detection method. *Analytical Chemistry* 2024, **96**(5)**:** 2107–2116.

34. Zhuang X, Sun R, Zhou D, Liu S, Wu Y, Shi Z*, et al.* Synergistic effects of multifunctional lanthanides doped CsPbBrCl2 quantum dots for efficient and stable MAPbI3 perovskite solar cells. *Advanced Functional Materials* 2022, **32**(18)**:** 2110346.

35. Liang L, Wang C, Chen J, Wang QJ, Liu X. Incoherent broadband mid-infrared detection with lanthanide nanotransducers. *Nature Photonics* 2022, **16**(10)**:** 712–717.

36. Verhagen E, Kuipers L, Polman A. Field enhancement in metallic subwavelength aperture arrays probed by erbium upconversion luminescence. *Optics express* 2009, **17**(17)**:** 14586–14598.

37. Zhang W, Ding F, Chou SY. Large enhancement of upconversion luminescence of NaYF4: Yb 3+/Er3+ Nanocrystal by 3D plasmonic nano-antennas. *Advanced Materials* 2012, **24**(35)**:** OP236–OP241.

38. Greybush NJ, Saboktakin M, Ye X, Della Giovampaola C, Oh SJ, Berry NE*, et al.* Plasmon-enhanced upconversion luminescence in single nanophosphor–nanorod heterodimers formed through template-assisted self-assembly. *ACS nano* 2014, **8**(9)**:** 9482–9491.

39. Luu Q, Hor A, Fisher J, Anderson RB, Liu S, Luk T-S*, et al.* Two-color surface plasmon polariton enhanced upconversion in NaYF4: Yb: Tm nanoparticles on Au nanopillar arrays. *The Journal of Physical Chemistry C* 2014, **118**(6)**:** 3251–3257.

40. Wu DM, García-Etxarri A, Salleo A, Dionne JA. Plasmon-enhanced upconversion. *The journal of physical chemistry letters* 2014, **5**(22)**:** 4020–4031.

41. Chen W, Roelli P, Hu H, Verlekar S, Amirtharaj SP, Barreda AI*, et al.* Continuous-wave frequency upconversion with a molecular optomechanical nanocavity. *Science* 2021, **374**(6572)**:** 1264–1267.

42. Xomalis A, Zheng X, Chikkaraddy R, Koczor-Benda Z, Miele E, Rosta E*, et al.* Detecting mid-infrared light by molecular frequency upconversion in dual-wavelength nanoantennas. *Science* 2021, **374**(6572)**:** 1268–1271.

43. Knight MW, Sobhani H, Nordlander P, Halas NJ. Photodetection with active optical antennas. *Science* 2011, **332**(6030)**:** 702–704.

44. Grajower M, Desiatov B, Mazurski N, Shappir J, Khurgin JB, Levy U. Optimization and experimental demonstration of plasmonic enhanced internal photoemission silicon Schottky detectors in the mid-IR. *Acs Photonics* 2017, **4**(4)**:** 1015–1020.

45. Su Z-C, Li Y-H, Lin C-F. Mid-infrared response from Cr/n-Si Schottky junction with an ultra-thin Cr metal. *Nanomaterials* 2022, **12**(10)**:** 1750.

46. Abadi A, Abubakr E, Oshita M, Noda D, Ohta R, Kan T. Characterization of an uncooled mid-infrared Schottky photodetector based on a 2D Au/PtSi/p-Si nanohole array at a higher light modulation frequency. *Applied Optics* 2024, **63**(8)**:** 2046–2055.

47. Shiraishi M, Noda D, Ohta R, Kan T. Room-temperature plasmonic mid-infrared photodetector based on PtSi/p-Si low Schottky-barrier junction. *Applied Optics* 2022, **61**(14)**:** 3987–3996.

48. Mehrfar AH, Majd AE. Responsivity enhancement of a PtSi photodetector with graphene by the photogating effect. *Applied Optics* 2023, **62**(5)**:** 1160–1166.

49. Yasunaga S, Kan T. Plasmonic mid-infrared photodetector with narrow trenches for reconstructive spectroscopy. *Optics Express* 2022, **30**(12)**:** 20820–20838.

50. Huang Y, Liang H, Zhang Y, Yin S, Li X, Cai C*, et al.* A high-performance broadband double-junction photodetector based on silicon nanowire arrays wrapped by silver nanoparticles for low-light imaging. *Journal of Materials Chemistry C* 2023, **11**(26)**:** 8784–8795.

51. Liu X, Zhou Q, Luo S, Du H, Cao Z, Peng X*, et al.* Infrared photodetector based on the photothermionic effect of graphene-nanowall/silicon heterojunction. *ACS applied materials & interfaces* 2019, **11**(19)**:** 17663–17669.

52. Sun R-L, Lai H-H, Su Z-C, Dong Y-H, Chen B-H, Sinha D*, et al.* Broad spectral response to photon energy unlimited by Schottky barrier from NiSi/Si junction. *Optics & Laser Technology* 2023, **166:** 109613.

53. Sharma KH, Dong Y-H, Chiang P-H, Su Z-C, Lin C-F. Silicon (100) surface passivation-driven tuning of Ag film crystallinity and its impact on the performance of Ag/n-Si mid-infrared Schottky photodetector. *APL Materials* 2024, **12**(6).

54. Su Z-C, Dong Y-H, Lin C-F. Hot-carrier effect and nanometer metal enabling Si-based mid-infrared detection beyond 5μm of wavelength. *IEEE Photonics Journal* 2024, **16**(3)**:** 1–10.

55. Wang CW, Liang L, Zhang X, Dai M, Chen J, Liu X*, et al.* Mid-infrared detection through ligand-driven local heating in lanthanide-doped nanoparticles. *Nature Communications* 2026.

56. Hornbeck J, Haynes J. Trapping of minority carriers in silicon. I. P-type silicon. *Physical Review* 1955, **97**(2)**:** 311.

57. Kaiser W, Frisch H, Reiss H. Mechanism of the formation of donor states in heat-treated silicon. *Physical Review* 1958, **112**(5)**:** 1546.

58. Hall R. Recombination processes in semiconductors. *Proceedings of the IEE-Part B: Electronic and Communication Engineering* 1959, **106**(17S)**:** 923–931.

59. Campabadal F, Milian V, Aymerich-Humet X. Trap-Assisted Tunneling in MIS and Schottky Structures. *physica status solidi (a)* 1983, **79**(1)**:** 223–236.

60. Tak T, Ho WY, Celupica-Liu I, Chow YC, Peretti J, Weisbuch C*, et al.* Trap-assisted Auger-Meitner recombination in GaN pin diodes. *Physical Review B* 2025, **111**(15)**:** 155308.